# A Noise-level-aware Framework for PET Image Denoising

Ye Li[1], Jianan Cui[1], Junyu Chen[2], Guodong Zeng[4], Scott Wollenweber[3], Floris Jansen[3], Se-In Jang[1], Kyungsang Kim[1], Kuang Gong[1] and Quanzheng Li[1]

[1] Center of Advanced Medical Computing and Analysis, MGH/HMS, Boston, MA 02155, USA
[2] Department of Electrical and Computer Engineering, Johns Hopkins University, Baltimore, MD 21218, USA
[3] GE Healthcare, Waukesha, WI, 53188, USA
[4] University of Bern, 3012 Bern, Switzerland
gary.li@mgh.harvard.edu

**Abstract.**
In PET, the amount of relative (signal-dependent) noise present in different body regions can be significantly different and is inherently related to the number of counts present in that region. The number of counts in a region depends, in principle and among other factors, on the total administered activity, scanner sensitivity, image acquisition duration, radiopharmaceutical tracer uptake in the region, and patient local body morphometry surrounding the region. In theory, less amount of denoising operations is needed to denoise a high-count (low relative noise) image than images a low-count (high relative noise) image, and vice versa. The current deep-learning-based methods for PET image denoising are predominantly trained on image appearance only and have no special treatment for images of different noise levels. Our hypothesis is that by explicitly providing the local relative noise level of the input image to a deep convolutional neural network (DCNN), the DCNN can outperform itself trained on image appearance only. To this end, we propose a noise-level-aware framework denoising framework that allows embedding of local noise level into a DCNN. The proposed is trained and tested on 30 and 15 patient PET images acquired on a GE Discovery MI PET/CT system. Our experiments showed that the increases in both PSNR and SSIM from our backbone network with relative noise level embedding (NLE) versus the same network without NLE were statistically significant with p<0.001, and the proposed method significantly outperformed a strong baseline method by a large margin.

# 1  Introduction

In recent years, due to the fast development of deep convolutional neural networks (DCNN), we have witnessed rapid progress in DCNN-based PET image denoising, with the goal to reduce the patient's administered activity (AA) or shorten the image acquisition duration while maintaining sufficient diagnostic image quality. Many of these deep-learning-based methods for PET image denoising have achieved better performance than traditional image enhancement methods in recovering the PSNR and SSIM [1-10].

Depending on the radiotracer used and the pharmacokinetics, the activity uptake concentration can differ in different organs/parts of the same patient and other patients. These different radiotracer uptakes intrinsically lead to varying number of counts (coincidence events) being generated at different regions of the body. Besides the tracer uptakes being distributed differently within the patient body, the total amount of AA to the patient, the scan acquisition duration, the patient local body morphometry surrounding the region, and scanner sensitivity can also affect the number of counts received by the detector in that region. Together, these factors lead to very different noise levels in different areas of a patient's PET image and in images acquired for different patients, using different imaging protocols and on different scanners.

In principle, the amount of Poisson noise can be quantified by considering the coefficient of variation (COV), defined as the standard deviation divided by the mean, which describes the relative level of noise in a sinogram bin and is given by

$$COV = \frac{\sqrt{m}}{m} = \frac{1}{\sqrt{m}}, \tag{1}$$

which shows that Poisson noise, while growing in absolute terms with the signal, is relatively smaller at higher count levels. These different noise levels lead to drastically different appearances of the images. Furthermore, we observe that the ranking of the relative noise levels is likely correlated with the amount of denoising operations needed for each image pair shown in Fig. 1. Specifically, less amount of denoising is needed for images with high counts as compared to the images with low counts. However, most of the current DCNN-based PET denoising methods are trained on the whole image, consisting of a hodgepodge of noise levels, resulting in learned filters to contain a mix of features present in different noise levels.

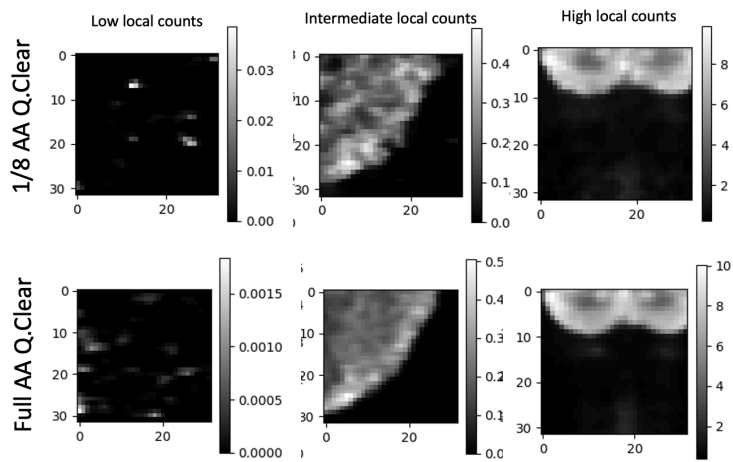

Fig. 1. Sample of coronal slices of patches cropped at three different regions of the same patient acquired with 1/8 AA and full AA. The images were reconstructed with GE's Q. Clear reconstruction algorithm and were normalized to SUV value by dividing (AA/weight). The scale bars indicate the range of counts present in each patch.

In this paper, we propose a noise-level-aware denoising framework that embeds a surrogate for the local noise level of the input image to condition the network to perform a denoising operation for a specified noise level. Specifically, the embedded noise level features enable the network to learn a specific set of filters needed for such specified noise-level mapping (from a high to low level of noise), as opposed to learning all the noise-level mappings altogether only from image appearance. The proposed framework was evaluated on two denoising tasks (1/8- and 1/4-to-full AA/acquisition duration) using a real patient dataset consisted of 45 images acquired with three different tracers (FDG, Fluciclovine and DOTATATE). The results show that a backbone network with noise level embedding (NLE) can outperform ($p<0.001$) the same network without NLE as measured by both the PSNR and SSIM.

## 2   Noise-level-aware Framework

As illustrated in Fig. 2, a surrogate scalar for the local relative noise level is embedded into a backbone denoising network to modulate the network for different denoising needs for input images with different relative noise levels. For the backbone network, we adopted the original resolution subnetwork (ORSNet) from the MPRNet [11]. The ORSNet does not employ any downsampling operation and generates spatially-enriched high-resolution features. The ORSNet is consisted of multiple original-resolution-blocks. Each of the ORB blocks further contains a channel-attenuation-block (CAB). A noise level embedding (NLE) layer is proposed to encode the noise level scalar to the CAB. The encoded feature vector is fed to multiple CABs to condition the importance of the feature maps on the relative noise level of the input image. The conditioning is done by first multiply the feature vector obtained from an adaptive max pooling layer with the first half of feature vector from the NLE layer and then add the resulting feature vector with the second half of feature vector from the NLE layer.

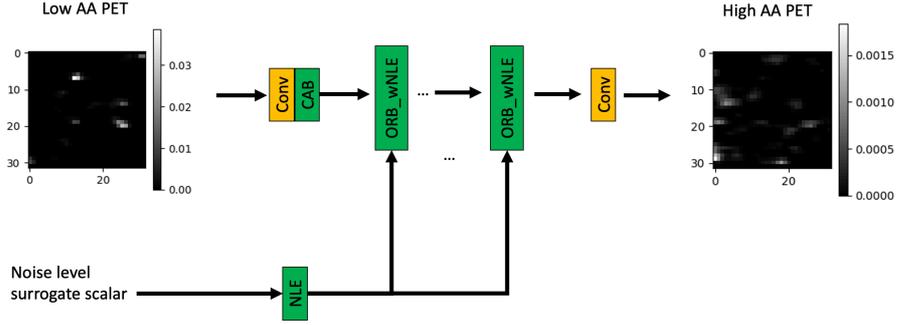

Fig. 2. Overall architecture of the proposed framework

The schematic of the proposed NLE layer and CAB block with noise level embedding are illustrated in Fig 3 (a) and (b), respectively.

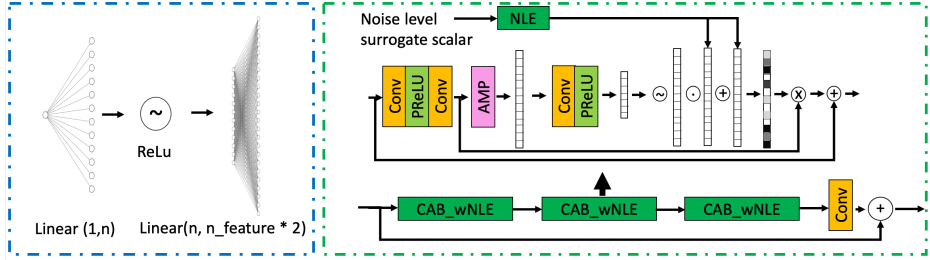

Fig. 3. The detailed architecture of the noise level embedding (NLE) block and modified channel attention block with NLE

## 2.1 Quantification of local relative noise level

In Fig. 1., we showed that images of different relative noise levels have very different appearances. This indicates that the filters learned from these images, if trained together with one network, may contain a mix of features present in these images, e.g., a feature that is present in images of one relative noise level can be learned together with a feature present in the images of another relative noise level. This would, in theory, lead to features learned images from a different relative noise level to emerge in one image. In attempt to solve this issue, we propose to split the image patches into groups that have similar appearances as categorized by the patches' relative noise level and its background. The rationale behind this is that we hypothesis the noise level embedding conditioned on each convolutional layer would help the network represent the importance of each feature map more accurately according to the input image's relative noise level, providing an extra aid to the DCNN in addition to its learning solely based on the image appearance.

To split the patches into groups, we applied Otsu's segmentation [12] on each patch, followed by computing the mean counts within the resulting 3D mask. The mean counts were used to quantify the local relative noise level present in each patch using equation 1. The patches were grouped into bins that fall in a range of relative noise levels. The ensembles of patches in these bins were expected to have similar noise level and image appearance. We used 4 bins with widths determined by visual inspection of the patches. The bins represent four types of patches: (1) high relative noise with clean background, (2) low relative noise with clean background, (3) high relative noise with lumpy background, and (4) low relative noise with lumpy background.

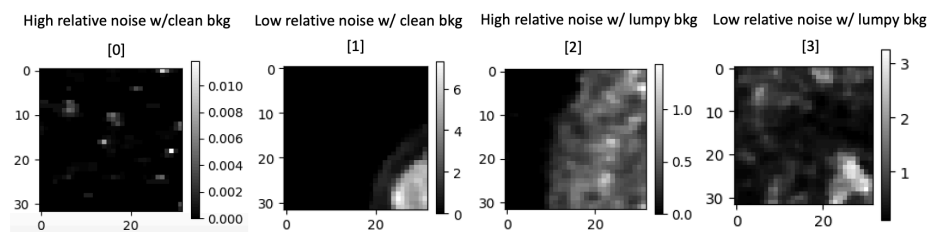

Fig. 4. Samples of patches that have different relative noise levels and image appearances.

## 3 Experiments

### 3.1 Experimental Setup

**Datasets**. A multi-tracer PET/CT dataset containing 45 patients was used to evaluate the proposed method. The dataset was split into a train and test dataset of 30 and 15 images, respectively. The range of administered activity of the dataset is 372 to 559 MBq. All patients were scanned with a GE Discovery MI 5-ring PET/CT system. 1/8 AA images were generated using listmode data of the full AA image. The random scatter correctors were generated based on the downsampled listmode data. PET images (matrix size: 352×224×128; voxel dimension: 2.8×2.734×2.734 2 mm$^3$) were reconstructed using GE's Q. Clear iterative reconstruction algorithm with beta of 750 for the full AA images and beta of 550 for the 1/8 AA images. Both the 1/8 AA and full AA images were normalized to the standard uptake value (SUV) by dividing the corresponding AA over the patient's weight.

### 3.2 Implementation Details

The proposed framework is end-to-end trainable and requires no pre-training. The network was trained on 32 × 32 × 32 patches with a batch size of 16 for 100 iterations. For data augmentation, horizontal and vertical flips are randomly applied. We used

Adam [13] optimizer with initial learning rate of $1 \times 10^{-5}$, which is steadily decreased to $1 \times 10^{-6}$ using the cosine annealing strategy [14].

### 3.3 Results and Analysis

**Ablation Study**. To examine the contribution of the NLE module, we conducted an ablation study using the same backbone network with and without NLE. The same train and test dataset was used to train and test these two networks. For the backbone network, we used the ORN as described in section 2. With the aid from the relative noise level embedding conditioned on each convolution operation, we expect that the feature maps to be modulated for the corresponding relative noise level of the image. This will enable the filters of the next convolutional layer to learn appropriate features (i.e., scattered points, sharp edges, and bright spots in lumpy background, etc.) for the images of that relative noise level.

**Table 1.** Ablation study on the 1/8 AA test dataset. ORN and NLE represent the original resolution network in [11], and noise level embedding proposed, respectively.

| Test Image Number | PSNR (input) | PSNR ORN only | PSNR ORNw/NLE | SSIM (input) | SSIM ORN only | SSIM ORNw/NLE |
|---|---|---|---|---|---|---|
| 59 | 41.49 | 44.17 | 44.75 | 0.88 | 0.93 | 0.93 |
| 66 | 43.67 | 46.03 | 46.6 | 0.89 | 0.94 | 0.94 |
| 67 | 43.93 | 45.9 | 46.05 | 0.88 | 0.93 | 0.93 |
| 68 | 52.79 | 54.68 | 54.83 | 0.9 | 0.95 | 0.95 |
| 69 | 53.64 | 54.8 | 55.06 | 0.93 | 0.96 | 0.96 |
| 70 | 40.44 | 43.53 | 43.79 | 0.9 | 0.94 | 0.95 |
| 71 | 42.62 | 44.95 | 45.46 | 0.87 | 0.93 | 0.93 |
| 72 | 57.62 | 58.45 | 58.92 | 0.92 | 0.95 | 0.96 |
| 73 | 57.61 | 59.44 | 59.69 | 0.89 | 0.94 | 0.94 |
| 74 | 47.43 | 48.67 | 48.89 | 0.88 | 0.93 | 0.93 |
| 77 | 49.8 | 51.4 | 51.95 | 0.93 | 0.96 | 0.96 |
| 78 | 53.52 | 55.91 | 56.01 | 0.9 | 0.94 | 0.95 |
| 80 | 55.9 | 56.64 | 57.15 | 0.91 | 0.95 | 0.95 |
| 81 | 41.36 | 44.84 | 45.05 | 0.85 | 0.92 | 0.93 |
| 84 | 42.38 | 45.12 | 45.18 | 0.87 | 0.92 | 0.93 |
| Mean | 48.28 | 50.302 | 50.625 | 0.893 | 0.939 | 0.943 |

A paired t-test was conducted to verify the statistical significance of the performance gain of the proposed method (ORN w/ NLE) versus ORN only. The p-values were 9.7e-6 and 9.6e-3 for PSNR and SSIM, respectively. To visualize the performance gain from the NLE module, we show a few inferenced patches from the ORN network trained with and without the NLE module. We observed that the improvements are more significant for patches of high and intermediate relative noise levels than those of low relative noise levels. This observation agrees with our hypothesis that patches of low relative noise level (high counts) needs less amount of denoising operations. The PSNR and SSIM improvements are likely attributed to the NLE module's ability to recover the noise-level-specific details shown in Fig. 5. Specifically, the noise-level-specific details missed by the backbone network are highlighted in the red circle and those recovered with addition of the NLE module are highlighted in the green circle.

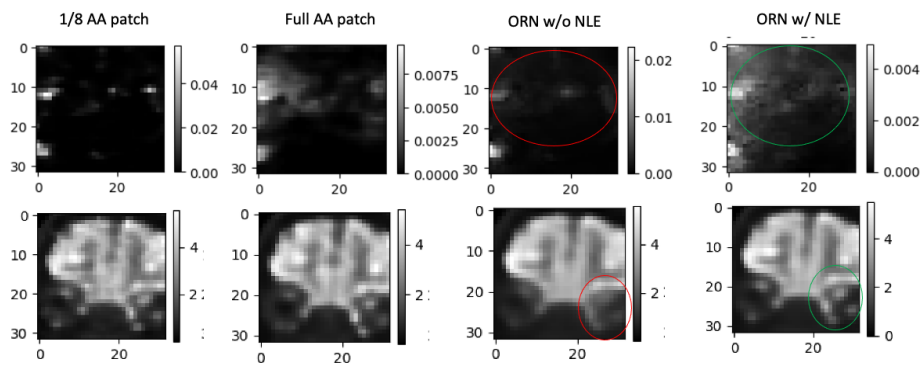

Fig. 5. Qualitative comparisons of sample testing slices with and without NLE. From left to right, 1/8 AA patch (input), Full AA patch (ground truth), predicted image without and with NLE. The scale bars indicate the ranges of voxel values present in the patches

**Comparison with State-of-the-Art** We compared the proposed method with a state-of-the-art image denoising method (MPRNet). To make the comparison fair, we modified the original 2D MRPNet to a 3D network. Furthermore, we adopted a strong baseline method as described in [4], which has been shown to outperform several reference methods commonly used in PET image denoising such as the Gaussian, NLM, BM4D, and Deep Decoder.

Fig. 6 summarizes the quantitative performance gains of the baseline, SOTA method and the proposed method. In addition, we quantified the performance gain by $\Delta$PSNR and $\Delta$SSIM between the proposed and baseline (orange) method. The results show that the 95% CI of the $\Delta$PSNR and $\Delta$SSIM between the proposed and baseline method, are [1.04, 2.26], [0.015, 0.025] and [0.916, 2.180] [0.009, 0.019], for the 1/8- and 1/4-full AA denoising task, respectively, under the assumption that the $\Delta$PSNR and $\Delta$SSIM were normally distributed. Fig. 7 shows sample images generated using the baseline and proposed method.

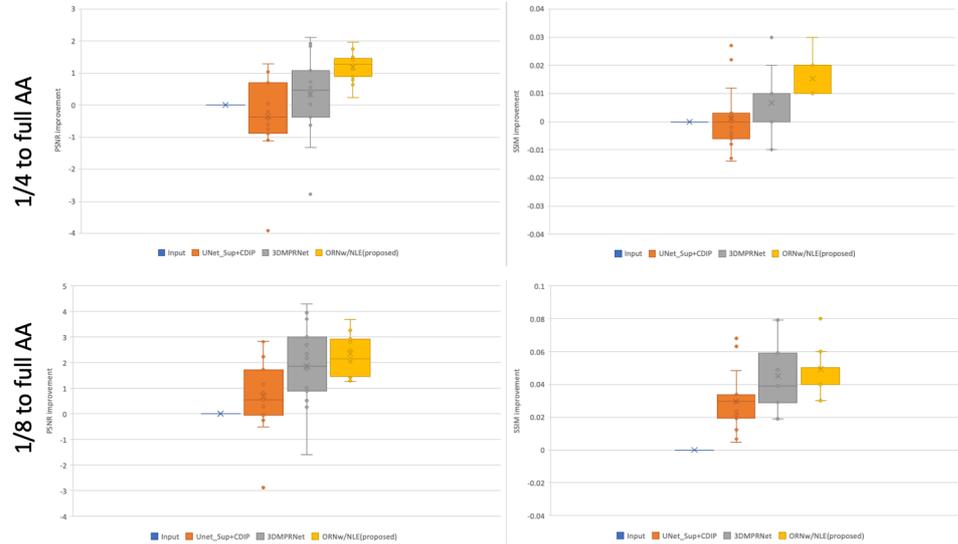

Fig. 6. Box plot of PSNR and SSIM improvement of the baseline method, 3DMPRNet, and proposed method.

## 4 Conclusion

The relative noise level embedding framework developed in this paper was able to outperform the baseline method for the task of PET image denoising as measured by SSIM and PSNR. This indicates that embedding relative noise level into a denoising network may help assist the network in recognizing and subsequently processing patterns that are specific to a pre-specified relative noise level. The proposed framework could be readily adapted to perform denoising task for other nuclear medicine imaging modality such as SPECT.

9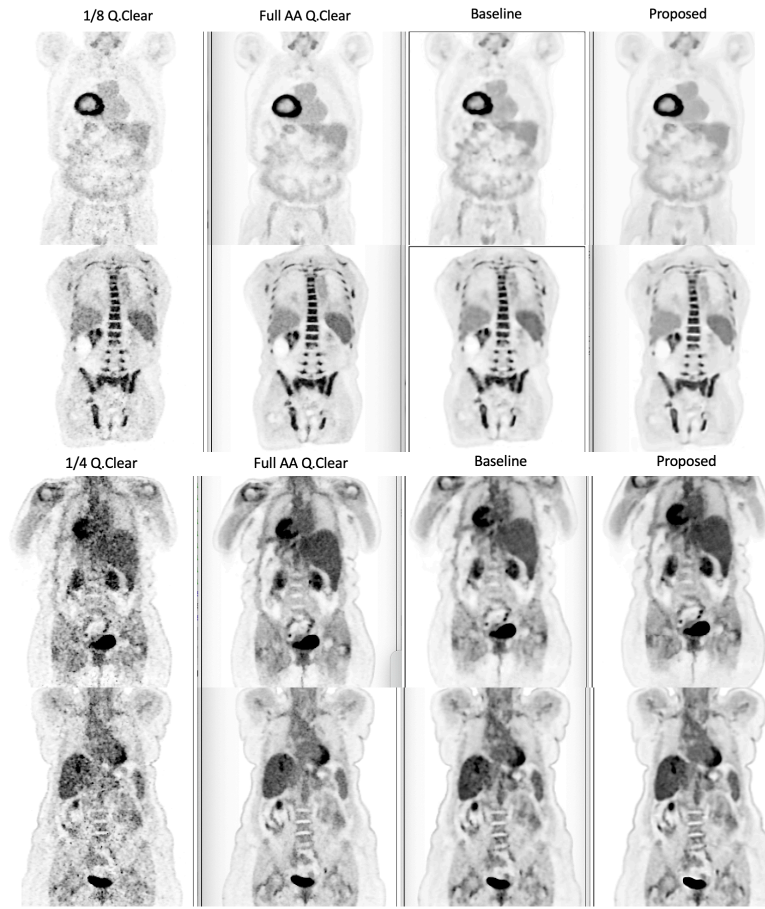

Fig. 7. Sample coronal slices of the 1/8, 1/4, and full AA Q. Clear reconstruction (ground truth), and predicted (on test data) image using the baseline and proposed method.

**References**

1. Gong, K., et al., *PET Image Denoising Using a Deep Neural Network Through Fine Tuning.* Ieee Transactions on Radiation and Plasma Medical Sciences, 2019. **3**(2): p. 153-161.
2. Dutta, J., R.M. Leahy, and Q. Li, *Non-local means denoising of dynamic PET images.* PLoS One, 2013. **8**(12): p. e81390.
3. Chan, C., et al., *Postreconstruction nonlocal means filtering of whole-body PET with an anatomical prior.* IEEE Trans Med Imaging, 2014. **33**(3): p. 636-50.

4. Cui, J.N., et al., *PET image denoising using unsupervised deep learning.* European Journal of Nuclear Medicine and Molecular Imaging, 2019. **46**(13): p. 2780-2789.
5. Ouyang, J.H., et al., *Ultra-low-dose PET reconstruction using generative adversarial network with feature matching and task-specific perceptual loss.* Medical Physics, 2019. **46**(8): p. 3555-3564.
6. Cui, J.A., et al., *Populational and individual information based PET image denoising using conditional unsupervised learning.* Physics in Medicine and Biology, 2021. **66**(15).
7. Zhou, L., et al., *Supervised learning with cyclegan for low-dose FDG PET image denoising.* Medical Image Analysis, 2020. **65**.
8. Zhou, B., et al., *MDPET: A Unified Motion Correction and Denoising Adversarial Network for Low-Dose Gated PET.* Ieee Transactions on Medical Imaging, 2021. **40**(11): p. 3154-3164.
9. Song, T.A., F. Yang, and J. Dutta, *Noise2Void: unsupervised denoising of PET images.* Physics in Medicine and Biology, 2021. **66**(21).
10. Onishi, Y., et al., *Anatomical-guided attention enhances unsupervised PET image denoising performance.* Medical Image Analysis, 2021. **74**.
11. Syed Waqas Zamir* 1 Aditya Arora* 1 Salman Khan2 Munawar Hayat3 Fahad Shahbaz Khan2 Ming-Hsuan Yang4, 6 Ling Shao1,2. *Multi-Stage Progressive Image Restoration*. in *CVPR*. 2021.
12. Otsu, N., *A Threshold Selection Method from Gray-Level Histograms.* IEEE Transactions on Systems, Man, and Cybernetics, 1979. **9**(1).
13. Ba, D.P.K.a.J., *Adam: A method for stochastic optimization.* 2014: arXiv:1412.6980.
14. Hutter, I.L.a.F., *SGDR: Stochastic gra- dient descent with warm restarts*, in *ICLR*. 2017.